\documentclass[aps,onecolumn,showkeys,showpacs,11pt]{revtex4}

\usepackage{graphicx,epic,eepic,epsfig,amsmath,latexsym,amssymb,verbatim,subfigure}

\input{Qcircuit}

\usepackage{theorem}
\newtheorem{definition}{Definition}

\newtheorem{lemma}{Lemma}

\newtheorem{fact}{Fact}
\newtheorem{theorem}{Theorem}

\def\squareforqed{\hbox{\rlap{$\sqcap$}$\sqcup$}}
\def\qed{\ifmmode\squareforqed\else{\unskip\nobreak\hfil
\penalty50\hskip1em\null\nobreak\hfil\squareforqed
\parfillskip=0pt\finalhyphendemerits=0\endgraf}\fi}
\def\endenv{\ifmmode\;\else{\unskip\nobreak\hfil
\penalty50\hskip1em\null\nobreak\hfil\;
\parfillskip=0pt\finalhyphendemerits=0\endgraf}\fi}

\mathchardef\ordinarycolon\mathcode`\:
\mathcode`\:=\string"8000
\def\vcentcolon{\mathrel{\mathop\ordinarycolon}}
\begingroup \catcode`\:=\active
  \lowercase{\endgroup
  \let :\vcentcolon
  }

\newcommand{\nc}{\newcommand}
\nc{\rnc}{\renewcommand}
\nc{\beq}{\begin{equation}}
\nc{\eeq}{{\end{equation}}}
\nc{\beqa}{\begin{eqnarray}}
\nc{\eeqa}{\end{eqnarray}}
\nc{\lbar}[1]{\overline{#1}}
\rnc{\bra}[1]{\langle#1|}
\rnc{\ket}[1]{|#1\rangle}
\nc{\ketbra}[2]{|#1\rangle\!\langle#2|}
\nc{\braket}[2]{\langle#1|#2\rangle}
\nc{\proj}[1]{| #1\rangle\!\langle #1 |}
\nc{\avg}[1]{\langle#1\rangle}
\rnc{\max}{\operatorname{max}}
\nc{\Rank}{\operatorname{Rank}}
\nc{\smfrac}[2]{\mbox{$\frac{#1}{#2}$}}
\nc{\Tr}{\operatorname{Tr}}
\nc{\ox}{\otimes}
\nc{\dg}{\dagger}
\nc{\dn}{\downarrow}
\nc{\cA}{{\cal A}}
\nc{\cB}{{\cal B}}
\nc{\cC}{{\cal C}}
\nc{\cD}{{\cal D}}
\nc{\cE}{{\cal E}}
\nc{\cF}{{\cal F}}
\nc{\cG}{{\cal G}}
\nc{\cH}{{\cal H}}
\nc{\cI}{{\cal I}}
\nc{\cJ}{{\cal J}}
\nc{\cK}{{\cal K}}
\nc{\cL}{{\cal L}}
\nc{\cM}{{\cal M}}
\nc{\cN}{{\cal N}}
\nc{\cO}{{\cal O}}
\nc{\cP}{{\cal P}}
\nc{\cR}{{\cal R}}
\nc{\cS}{{\cal S}}
\nc{\cT}{{\cal T}}
\nc{\cX}{{\cal X}}
\nc{\cZ}{{\cal Z}}
\nc{\csupp}{{\operatorname{csupp}}}
\nc{\qsupp}{{\operatorname{qsupp}}}
\nc{\var}{\operatorname{var}}
\nc{\rar}{\rightarrow}
\nc{\lrar}{\longrightarrow}
\nc{\polylog}{\operatorname{polylog}}
\nc{\eq}[1]{Eq.~(\ref{#1})}

\nc{\glneq}{~{\raisebox{0.6ex}{$>$}  \hspace*{-1.8ex} \raisebox{-0.6ex}{$<$}}~}
\nc{\gleq}{~{\raisebox{0.6ex}{$\geq$}\hspace*{-1.8ex} \raisebox{-0.6ex}{$\leq$}}~}
\def\a{\alpha}
\def\b{\beta}

\def\d{\delta}
\def\e{\epsilon}

\def\ph{\varphi}

\def\ps{\psi}

\def\D{\Delta}

\nc{\RR}{{{\mathbb R}}}
\nc{\CC}{{{\mathbb C}}}
\nc{\FF}{{{\mathbb F}}}
 \nc{\NN}{{{\mathbb N}}}
\nc{\ZZ}{{{\mathbb Z}}}
\nc{\PP}{{{\mathbb P}}}
\nc{\QQ}{{{\mathbb Q}}}
\nc{\UU}{{{\mathbb U}}}
\nc{\EE}{{{\mathbb E}}}
\nc{\id}{{\mathbb I}}

\nc{\be}{\begin{equation}}
\nc{\ee}{{\end{equation}}}
\nc{\bea}{\begin{eqnarray}}
\nc{\eea}{\end{eqnarray}}
\nc{\<}{\langle}
\rnc{\>}{\rangle}
\nc{\Hom}[2]{\mbox{Hom}(\CC^{#1},\CC^{#2})}
\nc{\rU}{\mbox{U}}

\nc{\ob}[1]{#1}

\begin{document}

\title{Multiparty data hiding of quantum information}

\author{Patrick Hayden}
\email{patrick@cs.caltech.edu}
\affiliation{Institute for Quantum Information, Caltech 107--81,
    Pasadena, CA 91125, USA}

\author{Debbie Leung}
\email{wcleung@cs.caltech.edu}
\affiliation{Institute for Quantum Information, Caltech 107--81,
    Pasadena, CA 91125, USA}

\author{Graeme Smith}
\email{graeme@theory.caltech.edu}
\affiliation{Institute for Quantum Information, Caltech 107--81,
    Pasadena, CA 91125, USA}

\begin{abstract}
We present protocols for multiparty data hiding of quantum information
that implement all possible threshold access structures. Closely related to
secret sharing, data hiding has a more demanding security requirement:
that the data remain secure against unrestricted LOCC attacks.
In the limit of hiding a large amount of data, our protocols
achieve an asymptotic rate of one hidden qubit per local physical
qubit.  That is, each party holds a share that is the same
size as the hidden state to leading order, with accuracy and security
parameters incurring an overhead that is asymptotically negligible.  
The data hiding states have very unusual entanglement properties,
which we briefly discuss.
%We end with a brief 
%discussion of the relationship between these protocols and the theory of 
%multipartite entanglement.
\end{abstract}

\date{July 14, 2004}
\maketitle

%%%%%%%%%%%%%%%%%%%%%%%%%%%%%%%%%%%%%%%%%%%%%%%%%%%%%%%%%%%%%%%%%%%%%
\section{Introduction} \label{sec:intro}

%In a variety of situations, it is desirable to encode data in a
%multipartite system in such a way that only certain authorized
%communication structures can be used to get access to that data.  
In a variety of situations, it is desirable to distribute data
among many parties in such a way that the parties can reconstruct
the data only if they cooperate in a well-defined way. This
problem has been studied in several settings, including the purely
classical case \cite{Sha79}, encoding classical data in quantum
systems \cite{TDL01,DLT02,EW02,EW04}, and encoding quantum data in
quantum systems~\cite{CGL99,G99,DHT02,HLSW03}.  
For the last two settings, at least
two inequivalent security criteria can be applied. In \emph{quantum
secret sharing} \cite{CGL99,G99}, certain authorized sets of parties are able
to reconstruct the encoded data if they cooperate to implement 
a joint operation, whereas the remaining, unauthorized, sets are 
unable to get access regardless of what
they do.    In
\emph{quantum data hiding} \cite{TDL01,DLT02,EW02,EW04,DHT02,HLSW03},
the requirement for cooperation is increased. The data must remain
inaccessible if any combination of the parties communicate classically,
and can only be retrieved if the members of an authorized set perform a joint
\emph{quantum mechanical} operation, perhaps supplemented by classical advice
from other parties outside the authorized set.

Consider the following fanciful scenario. After the debacle of the year 
2000 election, authorities in the state of Florida have decided to (do their
best to) implement a tamper-proof vote counting system for the upcoming
2004 election. One feature of the new system is that 
every counting center must be attended by both a Democratic
and a Republican observer. The system designers enforce this by encoding the
ballot box access code using a quantum data hiding scheme requiring
that the Democratic and Republican observers 
jointly implement a quantum operation
to get access to the code. The difficulty inherent in implementing a 
quantum operation remotely offers some assurance that both observers
will need to work together in the same place to extract the
code. The scheme also offers flexibility in incorporating smaller political
parties unable to field a full team of observers. The system could be
designed, for example, such that the Greens would hold a share of the
encoded state but not need to be physically present; their participation
via classical communication would be sufficient.

Relatively few data hiding schemes have been presented in the
literature.  The first examples are to be found in \cite{TDL01} and 
\cite{DLT02}, which demonstrate that classical bits can be hidden in bipartite
Werner states.  Generalizations to the multiparty setting realizing
all sensible access structures are given by \cite{EW02,EW04}.  The
earliest schemes for hiding quantum data \cite{DHT02} are based on
hiding a classical key that encrypts the quantum data.  
Much more efficient schemes for hiding quantum data directly
were presented for two parties in \cite{HLSW03}.
Here we generalize the method of \cite{HLSW03} to
construct $(k,n)$ threshold hiding schemes, meaning that arbitrary
classical communication in addition to quantum communication among any
$k$ parties is authorized, that is,
sufficient to retrieve the data, but arbitrary
classical communication along with quantum communication among groups
of up to $k-1$ parties is unauthorized. 
%(Note that since all parties participate in
%the state reconstruction, the no-cloning theorem does not require $k >
%\frac{n}{2}$, as it does in the case of quantum secret sharing.)  
Our contribution extends the multipartite threshold access structure
results in \cite{EW02,EW04} to the domain of hiding quantum information,
achieving an
asymptotic rate of one hidden (logical) qubit per local physical
qubit.
Like the schemes in \cite{HLSW03}, those presented here 
are significant improvements over the earlier multipartite 
results~\cite{EW02,DHT02} when a large number of qubits or even bits
are to be hidden -- the accuracy and security requirements incur
additive (negligible) space overhead in our schemes and multiplicative
(nonnegligible) overhead in previous constructions.
Also, unlike quantum secret sharing, where the no-cloning theorem imposes
the restriction that the complement of an authorized set be unauthorized, 
any threshold value $k$ between $1$ and $n$ is possible.
% 
% wcl: the comparison rewritten -- efficiency appears on have similar 
% dependencies on the security parameters if one hides only one bit 
%

While presented in the language of cryptography, quantum data hiding
is equally well a platform for the study of nonlocality. Indeed, the
original proposal of~\cite{TDL01} was motivated by the discovery
of quantum nonlocality without entanglement, meaning
sets of orthogonal product states that could not be distinguished
by local operations and classical communication (LOCC) 
alone~\cite{BDFMRSSW99}. Since the original discovery,
considerable effort has been devoted to understanding the relationship
between local distinguishability and other types of nonlocality.
(See~\cite{GKRSS01,WH02,HSSH02,CL03,DMSST03} and references therein.)
This paper continues that effort, in the sense that we present whole
subspaces all of whose states are indistinguishable by LOCC but that
can nonetheless be reconstructed by some collective operations, which
are now more carefully prescribed than in the earlier work.
Also, while much of that earlier work on local indistinguishability is 
devoted to determining when a finite set of states cannot be 
perfectly distinguished by LOCC, our focus here is at the other extreme, 
on near-perfect indistinguishability for entire subspaces.
Moreover, in contrast to the emphasis on product states in~\cite{BDFMRSSW99},
the results of~\cite{HLW04} ensure that the states we choose here
are by no means separable (they have near-maximal entanglement of
formation) despite being LOCC indistinguishable from the maximally
mixed state.

\bigskip
\noindent {\bf Notation:~}
We use the following conventions throughout the paper.  $\log$ and
$\exp$ are always taken base $2$. Unless otherwise stated, a \emph{state}
can be pure or mixed.  The density operator $\proj{\ph}$ of the pure
state $\ket{\ph}$ will frequently be written simply as $\ph$. 
%Sums and
%labels, unless otherwise stated, begin at 1. 
$\UU(d)$ denotes the
unitary group on $\CC^d$, $\cB(\CC^d)$ the set of linear transformations
from $\CC^d$ to $\CC^d$ and $I$ the identity matrix.  
\emph{Physical operations} mapping
$d_1$-dimensional states to $d_2$-dimensional states are completely
positive trace-preserving (CPTP) maps from $\cB(\CC^{d_1})$ to
$\cB(\CC^{d_2})$.  $\|\cdot\|_1$ denotes the trace norm of a
matrix and the $1$-norm of a vector while
$\|\cdot\|_2$ denotes the Hilbert space norm. $\Pr(E)$ is used to
represent the probability of event $E$.

\pagebreak

%%%%%%%%%%%%%%%%%%%%%%%%%%%%%%%%%%%%%%%%%%%%%%%%%%%%%%%%%%%%%%%%%%%%%
\section{Definitions and results}
We begin with a formal definition of data hiding for quantum information:
\begin{definition}
A \emph{$(\delta, \epsilon,s,d^n)$-qubit hiding scheme with $(k,n)$
access structure} consists of a CPTP encoding map, $E : \cB(\CC^s)
\rightarrow \cB(\CC^{d^n})$ and a set of CPTP decoding maps 
$D^{(X)}:\CC^{d^n} \rightarrow \cB(\CC^s)$,
one for each set $X$ of $k$ parties, that can be implemented via
quantum communication among the parties in $X$ together with classical
communication among all parties.  
% 
% wcl: the domain of D^X seems to be \CC^{d^n} 
%  
% We write the decoding maps as $D^{(X)} : 
% \cB(\CC^{d^k} \cong X \subset \CC^{d^n}) \rightarrow \cB(\CC^s)$. 
The encoding and decodings must satisfy
\begin{enumerate}
\item \emph{(Security)} For all states $\ph_0$ and $\ph_1$ on $\CC^s$,
along with all measurements $L$ that can be implemented using arbitrary 
quantum communication within groupings of $k-1$ or fewer parties and
arbitrary classical communication between them,
\begin{equation} \label{eqn:securityDefn}
\| L(E(\ph_0)) - L(E(\ph_1)) \|_1 \leq \e.
\end{equation}
\item \emph{(Correctness)} For all $X$ and for all states $\ph$ on
$\CC^s$, $\| (D^{(X)} \circ E)(\ph) - \ph \|_1 \leq \d$.
\end{enumerate}
\end{definition}
Notice that for $k=1$ there are no unauthorized measurements, so that a 
qubit hiding scheme with 
$(1,n)$ access structure is simply a method for implementing a 
kind of distributed data storage.
Any one of the $n$ parties can recover the quantum data with the help of
only classical ``advice'' from the other $n-1$ parties. We'll
see that even in this simple setting, our approach will be considerably
more efficient than the na\"{i}ve constructions based on local storage and
teleportation or the more sophisticated proposals in~\cite{B01}.

The encoding map we will use is of the form
\begin{equation}
E(\ph)  = \frac{1}{r}\sum_{i=1}^r U_i \ph U_i^{\dg},
\label{Eq:Edef}
\end{equation}
where the $U_i$ are in $\UU(d^n)$ and we fix some inclusion $\CC^s
\cong S \subset \CC^{d^n}$ for the encoded subspace; 
this is the same type of map used for approximate
randomization and bipartite data hiding in \cite{HLSW03}. Our strategy
will be to show that if the $U_i$ are selected independently and each
according to the Haar measure on $\UU(d^n)$, then for suitable choices
of the parameters, $E$ will provide a good encoding with non-zero
probability (over the random $U_i$). 
Given a set of parties in quantum communication with each other, $X$,
and its complement, $W$, our decoding map $D^{(X)}$ consists of local
measurements by the parties in $W$, communication of the measurement
outcomes to the parties in $X$, followed by a recovery procedure on
$X$.
More specifically, the decoding is achieved in a two step process. In
the first step, $A^{(W)}$, each member of $W$ performs a projection
onto a fixed local basis, collectively $\{\ket{l}_W \}_{l=1}^{d^{n-k}}$, and 
sends the outcome to the members of $X$. Thus, if $W$ consists of the parties
$W_1, W_2, \ldots, W_{n-k}$, then $A^{(W)}$ has the structure
$A^{(W)} = A^{(W_1)} \ox A^{(W_2)} \ox \cdots \ox A^{(W_{n-k})}$.
An arbitrary hidden state $\ket{\ph} \in S$ is thereby transformed to the state
$A^{(W)}(E(\ph)) = \sum_{l} (\proj{l}_W \ox I_X) E(\ph) (\proj{l}_W
\ox I_X)$, now entirely in the possession of the members of $X$ since,
post-measurement,
the system $W$ can be assumed to contain only the measurement outcomes,
which get sent to $X$.

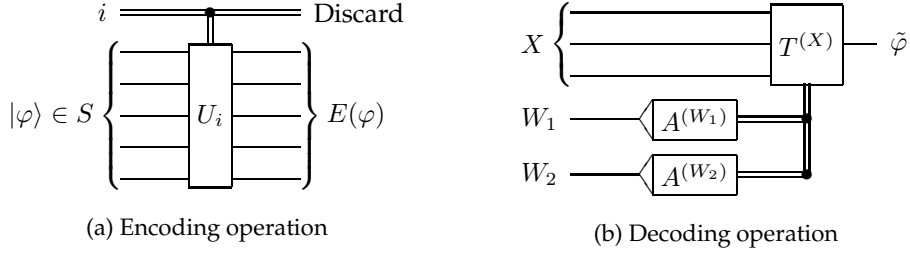
\begin{figure}
\subfigure[Encoding operation]{
\hspace{1.5cm}
\Qcircuit @C=1.3em @R=.6em {
\lstick{i} & \cw & \control \cw            & \cw & \rstick{\mbox{Discard}}\cw \\
           &     & \cwx                    &     &     \\
           & \qw & \multigate{4}{U_i} \cwx & \qw & \qw \\
           & \qw & \ghost{U_i}             & \qw & \qw \\
\lstick{\ket{\ph} \in S~~} 
           & \qw & \ghost{U_i}             & \qw & \rstick{~~E(\ph)} \qw \\
           & \qw & \ghost{U_i}             & \qw & \qw \\
           & \qw & \ghost{U_i}             & \qw & \qw 
\gategroup{3}{1}{7}{1}{.7em}{\{}
\gategroup{3}{5}{7}{5}{.7em}{\}}
} 
\hspace{1.5cm}
\label{subfig:1}
}
\subfigure[Decoding operation]{
\hspace{1.5cm}
\Qcircuit @C=1.3em @R=.6em {
             & \qw & \qw & \multigate{2}{T^{(X)}} & \\
\lstick{X~~} & \qw & \qw & \ghost{T^{(X)}}        & \rstick{\tilde{\ph}} \qw \\
             & \qw & \qw & \ghost{T^{(X)}}        & \\
\lstick{W_1} & \qw & \measuretab{A^{(W_1)}} & \control \cw \cwx \\
\lstick{W_2} & \qw & \measuretab{A^{(W_2)}} & \control \cw \cwx 
\gategroup{1}{1}{3}{1}{.7em}{\{}
} 
\hspace{1.5cm}
\label{subfig:2}
}
\caption{The encoding and decoding maps. \ref{subfig:1} 
depicts the encoding procedure. 
A random $U_i$ is applied to the state $\ket{\ph}$ drawn from 
subspace $S$. The output, $E(\ph)$, is almost indistinguishable 
from the maximally mixed state using arbitrary LOCC and quantum communication
within groupings of $k-1$ and fewer parties. \ref{subfig:2} depicts the
decoding procedure. For any partition of the parties into an authorized
set $X$ of $k$ parties and its complement $W$ of $n-k$ parties, the
unauthorized parties $W_1, W_2, \ldots, W_{n-k}$ perform the measurements
$A^{(W_j)}$ in fixed local bases, sending the outcomes
to the authorized parties, who then apply the
tranpose channel $T^{(X)}$. (Single lines represent quantum data, double lines
classical data. Time flows from left to right.)}
\label{fig:hide}
\end{figure}

They will then perform the transpose channel $T^{(X)}$
\cite{OhyaP93,Barnum} of the CPTP map $A^{(W)} \circ E$ adapted to the
maximally mixed state on $S$.  The transpose channel is a generic
construction for approximately reversing a quantum operation, which
in the present case leads to the following definition for $T^{(X)}$.
Let $P_l := \proj{l}_W \ox I_X$, $P_S$ be the projector onto $S$, and
\bea 
 N := \sum_{i=1}^s \sum_{l=1}^{d^{n-k}} P_l\frac{U_i}{\sqrt{r}} \frac{P_S}{s} 
     \frac{U_i^\dg}{\sqrt{r}} P_l \,.  
\label{eq:NDef}
\eea
Then, 
\begin{equation}\label{Eq:TransposeDef}
T^{(X)}(\psi) := \sum_{il} T_{il} \psi T_{il}^\dg \,, \quad % \hspace*{5ex}
{\rm where} \quad T_{il} :=
\frac{P_S}{\sqrt{s}}\frac{U_i^\dg}{\sqrt{r}}P_l N^{-1/2} \,.
\end{equation}
($N^{-1/2}$ is here defined to be zero outside the support of $N$. The 
map $T^{(X)}$ is, therefore, defined on the image of the subspace $S$ by
the map $A^{(W)} \circ E$ and can be extended to a CPTP map on all of
$W \ox X$.)
Formally, the decoding map is given by 
\bea
D^{(X)} = T^{(X)} \circ A^{(W)} \,.
\label{Eq:DecodingDef}
\eea
The arrangement is illustrated in Figure \ref{fig:hide}.
Our main result is
\begin{theorem} \label{thm:main}
Let $d$ be sufficiently large that $d^k > 48/\d^2$, $d^n > 10(n+2)/\e$
and, in the special case $k=1$, 
$\smfrac{d}{\log d} > \smfrac{2840(2n+3)}{\d^2}$. Then if
\begin{eqnarray}
 \nonumber 
r &=& \left\lceil 
      \frac{32(n+2)^4}{C \e^2} \cdot d^{k-1}\log d \right\rceil 
      \quad \mbox{and} \\
s &=& \left\lfloor 
      \frac{C\e^2\d^2}{1536(n+2)^4} \cdot \frac{d}{\log d} \right\rfloor,
\label{Eq:RandS1}
\end{eqnarray}
with probability at least $1/2$, the encoding map (\ref{Eq:Edef})
and decoding maps (\ref{Eq:DecodingDef}) give a
$(\delta,\epsilon,s,d^n)$-qubit hiding scheme with $(k,n)$ access
structure.
\end{theorem}

The intuition leading to these choices for $r$ and $s$ is quite simple. First,
security of the encoding against groupings of $k-1$ parties will require that
our encoding randomize subsystems of dimension $d^{k-1}$ and smaller. This
leads to a choice of $r \gg d^{k-1}$. On the other hand, successful decoding
will require the $k$ parties in $X$ to be able to identify the unitary $U_i$ 
that was applied to the input state $\ket{\ph} \in S$ without damaging the
encoded state, which leads to the constraint $r s \ll d^k$. Consequently,
in light of the randomization requirement, $s \ll d$.

\section{Proofs}

To prove the theorem, we will make extensive use of some well-known
facts about Gaussian random variables and random quantum states. We
use the notation $g \sim \cN_{\CC}(0,1)$ to denote that $g$ is a
complex Gaussian random variable with mean $0$ and variance $1$. That
is, $g = g^{(x)} + i g^{(y)}$ where $g^{(x)}$ and $g^{(y)}$ are
independent, mean $0$, variance $1/2$ \emph{real} Gaussian random
variables.
\begin{fact}\label{fact:1}
(Lemma 23 of \cite{BHLSW03})
Let $g_i \sim \cN_{\CC}(0,1)$ be independent complex Gaussian variables. 
Then, for $\e \geq 0$ the 
probabilities of large deviations are given by
% \begin{eqnarray}
% P\Big(\frac{1}{N} \sum_{i=1}^N |g_i|^2 > 1 + \e\Big) 
%  & \leq & \exp\Big(-N \frac{\e - \ln(1+\e)}{2\ln2}\Big) \quad \mbox{and}\\
% P\Big(\frac{1}{N} \sum_{i=1}^N |g_i|^2 < 1 - \e\Big) 
%  & \leq & \exp\Big(-N \frac{-\e - \ln(1-\e)}{2\ln2}\Big).
% \end{eqnarray}
% 
\begin{eqnarray}
\Pr\Big(\frac{1}{N} \sum_{i=1}^N |g_i|^2 
 \glneq
 1 \pm \e\Big) 
 \leq  
  \exp\Big({-}N \; \frac{\pm \e - \ln(1 \pm \e)}{\ln2}\Big) \,.
\end{eqnarray}
In particular, for $-1 \leq \d \leq 1$,
we have $ \d - \ln(1 + \d) \geq \frac{\d^2}{6}$, which implies 
that for $0 \leq \e \leq 1$,
\begin{equation} \label{Eq:epsilonsquared}
\Pr\Big(\frac{1}{N}\sum_{i=1}^N |g_i|^2 
 \glneq 
 (1 \pm \e)\Big) \leq  \exp({-}CN\e^2) \,,  
\end{equation}
where $C$ can be taken to be $(6\ln 2)^{-1}$.
\end{fact}
This can be used to derive:
\begin{fact}\label{fact:2}
(adapted from Lemma II.3 of \cite{HLSW03})
Let $\ph$ be a pure state and $P$ be a projector of rank $p$, both on a Hilbert space of dimension d.
If $\{U_i\}_{i=1}^N$ are chosen independently and
according to the Haar measure on $\UU(d)$, then
there exists a constant $C \geq (6\ln 2)^{-1}$ such that
\begin{equation}
\Pr\left ( \frac{1}{N} 
\sum_{i=1}^N\Tr(U_i\proj{\ph}U_i^\dg P) -\frac{p}{d}
 \glneq \pm\frac{\e p}{d} \right) 
 \leq  
  \exp\Big({-}Np \; \frac{\pm \e - \ln(1 \pm \e)}{\ln2}\Big) \,.
\end{equation}
If $0 \leq \e \leq 1$, we get the simpler upper bound
$\exp(-CNp\e^2)$ as in Fact~\ref{fact:1}.
\end{fact}
We will also use:
\begin{fact}\label{fact:3}
(Lemma II.4 of \cite{HLSW03}) For $0 < \e <1$ and dim $\cH = d$ there
exists a set $\cN$ of pure states in $\cH$ with $\left | \cN\right |
\leq (5/\e)^{2d}$ such that for every pure state $\ket{\ph} \in \cH$
there exists $\ket{\tilde{\ph}} \in \cN$ with $ \parallel \proj{\ph} -
\proj{\tilde{\ph}} \parallel_1 \leq \e$. (We call such a set an
\emph{$\e$-net}.)
\end{fact}

\subsection*{Theorem \ref{thm:main}: Proof of security}
Security is guaranteed if no unauthorized measurement can distinguish
any encoded state from the maximally mixed state.
We would like to show that the probability (over random choices of
$U_i$) of the contrary is small.
An unauthorized measurement is LOCC and thus separable~\cite{BNS98,BDFMRSSW99} 
over a partition of the $n$ parties into groups of size $< k$. We will
actually prove security against this larger class of measurements.
It suffices to consider measurements with rank one POVM elements since
any measurement can be refined to such a measurement without
decreasing distinguishability.
For example, a measurement implemented by LOCC between three groups 
of parties $W_1$, $W_2$ and $W_3$ will have a POVM of the form
$\{ Z_i = Z_i^{(1)} \ox Z_i^{(2)} \ox Z_i^{(3)} \}$, where each
$Z_i^{(j)}$ is an operator on the space $W_j$. Suppose that it is known
that
\begin{equation}
\Big| \Tr[ Z_i E(\ph) ] - \frac{\Tr[Z_i]}{d^n} \Big| 
\leq \frac{\e\Tr[Z_i]}{2d^n}
\end{equation}
for all states $\ket{\ph} \in S$. Then
\begin{equation}
\sum_i \Big| \Tr[ Z_i E(\ph) ] - \Tr[ Z_i \frac{I}{d^n} ] \Big| 
\leq \frac{\e}{2},
\end{equation}
confirming the security condition, Eq.~(\ref{eqn:securityDefn}), 
for this particular POVM by the triangle inequality.
Thus, it is sufficient to bound
\begin{equation}\label{eqn:securityBound}
\Pr\Big(\sup_{\ket{\ph} \in S} \sup_{Z} \Big|
    \Tr[\frac{1}{r} \sum_{i=1}^r ZU_i \ph U_i^{\dg}] - \frac{1}{d^n}
    \Big| \geq \frac{\e}{2d^n} \Big),
\end{equation}
where the second supremum is over all rank one projectors $Z$ of the
form $Z = \ox_{q=1}^nZ_{n_q}$, with $\sum_{q=1}^n n_q =n$, $0 \leq n_q
< k$ corresponding to some partition of the $n$ parties into
groups of size $n_q$ and each $Z_{n_q}$ a rank one projector 
on $\CC^{d^{n_q}}$. Now, if we let each $\cN_{n_q}$ be a
$\frac{\e}{2(n+2)d^n}$-net for states on $\CC^{d^{n_q}}$ and $\cN_S$ an
$\frac{\e}{2(n+2)d^n}$-net on $S$, then
\begin{equation}\begin{split}
 \Big|\Tr[\frac{1}{r} \sum_{i=1}^r
    (\ox_{q=1}^nZ_{n_q})  U_i \ph U_i^{\dg}] - \frac{1}{d^n}
 \Big| \geq \frac{\e}{2d^n}\Longrightarrow \\
 \Big|\Tr[\frac{1}{r} \sum_{i=1}^r
   (\ox_{q=1}^n \tilde{Z}_{n_q}) U_i \tilde{\ph} U_i^\dg]
        - \frac{1}{d^n} \Big| \geq\frac{\e}{2(n+2)d^n}
\end{split}\end{equation}
for some $\tilde{Z}_{n_q} \in \cN_{n_q}$, $\tilde{\ph} \in \cN_s$.  By
the union bound, the probability of Eq.~(\ref{eqn:securityBound}) is
therefore bounded above by
\begin{equation}\label{Eq:SecMeshed}
\underset{\{n_q \}}{\max} N_{\{n_q\}} 
\underset{\tilde{\ph}, \tilde{Z}}{\max}
    \Pr\Big(
     \Big|
    \Tr[\frac{1}{r} \sum_{i=1}^r\tilde{Z}U_i \tilde{\ph} U_i^{\dg}]
        - \frac{1}{d^n}
    \Big| \geq \frac{\e}{2(n+2)d^n}\Big),
\end{equation}
where $N_{\{n_q\}} = \frac{n!}{\prod n_q!}  (\prod_q
|\cN_{n_q}|)|\cN_s|$ is the number of different ways $n$ parties can
be divided into $n$ groups of size $n_q$, multiplied by the total
number of net points that need to be verified. In
Eq.~(\ref{Eq:SecMeshed}), the first maximization is taken over 
partitions $\{n_q\}$ 
of $n$ such that $0 \leq n_q < k$, while inner maximization
is taken over $\ket{\tilde{\ph}} \in \cN_s$
and $\tilde{Z} \in \times_q \cN_{n_q}$. (This is a slight abuse of
notation; note that $\tilde{Z}$ is a rank one projector on the
corresponding tensor product space.)  Applying Facts \ref{fact:2} and
\ref{fact:3}, this is bounded above by
\begin{equation}
 2^{n \log n}\Big(\frac{10(n+2)d^n}{\e}\Big)^{2(n+1)d^{k-1}}
 \exp\Big(\frac{-Cr\e^2}{4(n+2)^2}\Big),
\end{equation}
as long as $s \leq d^{k-1}$, which is the case for $k > 1$.  If $k=1$, 
all operations are authorized so there is no need for a security requirement.
If $d^n> \frac{10(n+2)}{\e}$, this is
less than $\frac{1}{4}$ for $r = \left\lceil \frac{32(n+2)^4}{C \e^2}
d^{k-1}\log d \right\rceil$.

\subsection*{Theorem \ref{thm:main}: Proof of correctness}

Our goal is to show that for any $X$ and for all $\ket{\ph} \in S$,
$\| D^{(X)} \circ E(\ph) - \ph \|_1 \leq \d$,
again with probability at least $3/4$ over choices of $\{U_i\}$,
so that the probability that the statement isn't true is no more than
$1/4$.  
This would be implied
by $\bra{\ph} (D^{(X)} \circ E)(\ph) \ket{\ph} \geq 1 - \d^2 / 4$ for
all $\ket{\ph} \in S$~\cite{FvG99}, which would in turn be implied by 
\bea
\label{eq:sufficient}
\forall i,l~~~
\frac{|\<\ph| T_{il} P_l U_i |\ph\>|^2}{\| P_l U_i |\ph\>\|_2^2} \geq
1 - \frac{\d^2}{4} := 1 - \a.
\eea
So, it suffices to show that Eq.~(\ref{eq:sufficient}) holds with
sufficiently high
probability for any fixed $\ket{\ph} \in S$ that it can be
achieved simultaneously for a net of states on $S$.

The proof idea is to take $\ket{\ph}$ as a member of an orthonormal 
basis of $S$, $\{ \ket{j} \}_{j=1}^s$.  
The success of using $T^{(X)}$ to decode the states $\{P_l U_i
\ket{j}\}_{i,j,l}$ can be gauged by considering it as the first piece
of a two-stage implementation of the Pretty Good Measurement
(PGM)~\cite{HausladenW94}, the criterion for success of which is
well-understood~\cite{HausladenJSWW96}.
More specifically, let $\ket{\xi_{ijl}} = P_l \frac{U_i}{\sqrt{r}}
\frac{\ket{j}}{\sqrt{s}}$, which is non-zero with probability one.  According to Eq.~(\ref{eq:NDef}), $N =
\sum_{ijl}\proj{\xi_{ijl}}$.  For any state $\rho$,
\begin{eqnarray}
	\bra{j}T_{il} \, \rho \, T_{il}^\dg \ket{j} 
& = & 	\Tr[\rho \, T_{il}^\dg \proj{j}T_{il}]
\label{eq:pgm1}
\\
& = & 	\Tr\Big[\rho \, N^{-1/2}P_l \frac{U_i}{\sqrt{r}} \frac{\proj{j}}{s} 
	\frac{U_i^\dg}{\sqrt{r}}P_l N^{-1/2} \Big] 
\\
& = & \Tr[\rho \, N^{-1/2}\proj{\xi_{ijl}}N^{-1/2}]
\\
& = & \Tr[\rho \, M_{ijl}],
\label{eq:pgm4}
\end{eqnarray}
where $M_{ijl} := N^{-1/2}\proj{\xi_{ijl}}N^{-1/2}$ is a POVM
element of the PGM on the set of unnormalized states $\{\proj{\xi_{ijl}}\}$.  
Letting $\rho =\frac{P_l
U_i \proj{j} U_i^\dg P_l}{\<j|U_i^\dg P_l U_i |j\>}$, the LHS of
Eqs.~(\ref{eq:sufficient}) and (\ref{eq:pgm1}) coincide, while
Eq.~(\ref{eq:pgm4}) is the probability that the PGM will correctly
identify $\frac{P_l U_i |j\>}{\|P_l U_i |j\>\|_2}$.  That is,
\begin{equation}
  \frac{ |\bra{j} T_{il} P_l U_i \ket{j}|^2 }{\| P_l U_i |\ph\>\|_2^2}
= \frac{ \Tr[\proj{\xi_{ijl}} M_{ijl}] }{\| \ket{\xi_{ijl}} \|_2^2 } \,.
\end{equation}

We must therefore bound the probability of error for the PGM which,
it is important to observe, is defined
on a set of sub-normalized states, where the
normalization of each state gives its probability.
An error bound for any equiprobable ensemble is given in
\cite{HausladenJSWW96}, and we provide a straightforward
generalization to the present case of unequal a priori probabilities
in Appendix \ref{appendix:PGM}.
Applying this error bound, and using the orthogonality of the states
$\ket{\xi_{ijl}}$ for different values of $l$, we obtain the following.  
 
\begin{lemma}
% For each $i = 1,\ldots,r$ and $l = 1,\ldots,d^{n{-}k}$, and the hidden
% quantum state $\ket{j}$, the fidelity between $\ket{j}$ and the output
% of $D^{(X)}$ is upper-bounded by the probability that the Pretty Good
% Measurement correctly identifies $\frac{P_l U_i \ket{j}}{\|P_l U_i
% \ket{j}\|_2}$, minimized over $i,j,l$.
% 
For each $i = 1,\ldots,r$, ~$l = 1,\ldots,d^{n{-}k}$, and $j = 1,\ldots,s$, 
\begin{eqnarray} \label{Eq:defnDelta}
 1 - \frac{ |\<j| T_{il} P_l U_i |j\>|^2 }{\|P_l U_i |j\> \|_2^2} 
 \leq
 \D_{ijl} := \frac{1}{|\bra{j}U_i^\dg P_l U_i \ket{j}|^2}
 \sum_{(i^\prime,j^\prime) \neq (i,j)} |\bra{j^\prime}U_{i^\prime}^\dg
 P_l U_i \ket{j}|^2.
\end{eqnarray}
\end{lemma}
(Note that the sum ranges over values of $i'$ from $1$ to $r$ and
$j'$ from $1$ to $s$, not including the pair $(i,j)$.)
The intuition behind this result is clear -- our probability of 
misidentification for a fixed state scales roughly like 
the sum of the overlaps of that state with all the states 
we could mistake it for, divided by a normalization factor.

In order to evaluate $\Pr (\D_{ijl} > \a)$, and thus determine the 
probability (over random choices of $U_i$) that the probability of 
misidentifying $P_lU_i\ket{j}$ (in the PGM) is small, 
we break up the above sum into two terms:
\begin{eqnarray}
\D_{ijl} &= &\frac{1}{|\bra{j}U_i^\dg P_l U_i \ket{j}|^2}
 \sum_{j^\prime}\sum_{i^\prime \neq i} |\bra{j^\prime}U_{i^\prime}^\dg P_l U_i \ket{j}|^2 + \frac{1}{|\bra{j}U_i^\dg P_l U_i \ket{j}|^2}
 \sum_{j^\prime \neq j}|\bra{j^\prime}U_{i}^\dg P_l U_i \ket{j}|^2 \nonumber\\
 &=:& \D_{ijl}^1+\D_{ijl}^2.
\end{eqnarray}
We can use Fact \ref{fact:2} to control the size of the denominator in $\D_{ijl}$ with the result that 
\begin{equation} \label{Eq:denomUB}
\Pr \Biggl( \frac{1}{\bra{j}U_i^\dg P_l   U_i\ket{j}} \geq 2d^{n-k}\Biggr) 
\leq \exp\Big(-\frac{d^k}{24\ln 2 }\Big).
\end{equation}

In general, if a particular event $E$ is excluded by a set of
conditions $C_1 \wedge C_2 \wedge \cdots$, then $\Pr(E) \leq \Pr(\neg C_1)
+ \Pr(\neg C_2) + \cdots$.  (This holds for arbitrary dependence between
$E,C_1,C_2,\ldots$) We will use this observation repeatedly in the 
arguments below.

Turning our attention to $\D^1_{ijl}$, we see that 
\begin{eqnarray}
 \D_{ijl}^1& = & \frac{1}{|\bra{j}U_i^\dg P_l U_i \ket{j}|^2}
 \sum_{j^\prime}\sum_{i^\prime \neq i} |\bra{j^\prime}U_{i^\prime}^\dg P_l U_i \ket{j}|^2 \nonumber\\
& = & \frac{1}{\bra{j}U_i^\dg P_l U_i \ket{j} } \sum_{j^\prime}\sum_{i^\prime \neq i} 
\Tr\Bigg[\proj{j^\prime}U^\dg_{i^\prime} \frac{P_l U_i \proj{j} U_i^\dg P_l}{\bra{j}U_i^\dg P_l U_i\ket{j}} U_{i^\prime}\Bigg]
\end{eqnarray}
so, using Eq.~(\ref{Eq:denomUB}) we find that
$\Pr( \D^1_{ijl} > \beta )$ is bounded above by
\begin{equation}
\Pr \Bigl( 2 d^{n-k}\sum_{j^\prime}\sum_{i^\prime \neq i} 
\Tr[\proj{j^\prime}U^\dg_{i^\prime} \frac{P_l U_i \proj{j} U_i^\dg P_l}{\bra{j}U_i^\dg P_l U_i\ket{j}} U_{i^\prime}] > \beta\Bigr)
+ \exp\Big(-\frac{d^k}{24\ln 2 }\Big).
\end{equation}
If we choose $(r-1)s \leq \b d^k / 3$ and apply Fact \ref{fact:2}, 
we see that this is no greater than 
\begin{equation}\label{Eq:D1UB}
\exp\Big(-\frac{(r-1)s}{24\ln 2}\Big)
+ \exp\Big(-\frac{d^k}{24\ln 2 }\Big).
\end{equation}
We must also deal with $\D_{ijl}^2$.  For $k=n$, this is identically zero, whereas for $1 \leq k < n$ we rely on the following lemma, the proof of which can
be found in Appendix \ref{appendix:errorControl}.
\begin{lemma} \label{lem:errorControl}
If $0 < \b,\e \leq 1$, \, $s \leq \b d /128$ \, and $1 \leq k < n$,
then
\begin{equation}\label{Eq:D2UB}
\Pr \Biggl(\frac{1}{|\bra{j}U_i^\dg P_l U_i \ket{j}|^2} \sum_{j^\prime \neq j}|\bra{j^\prime}U_{i}^\dg P_l U_i \ket{j}|^2  > \beta \Biggr) 
\leq 4s \exp\Big(-\frac{\beta d^k}{128\ln2}\Big).
\end{equation}
\end{lemma}
The lemma applies if we choose $s = \left\lfloor\frac{\beta C\e^2}{96(n+2)^4}\frac{d}{\log d} \right\rfloor$ with $d$ large enough that $s > 1$.
Together, Eqs.~(\ref{Eq:D1UB}) and (\ref{Eq:D2UB}) 
imply (letting $\beta = \frac{\a}{4}$) that if we choose $r$ and $s$ 
such that $\a d^k / 21 \leq (r-1)s \leq {\a d^k}/{12}$, 
\begin{eqnarray}\nonumber
\Pr \Bigl(  \D_{ijl} > \frac{\a}{2}\Bigr) 
&\leq& 4s \exp\Big(-\frac{\a d^k}{512\ln2}\Big) 
       + \exp\Big(-\frac{(r-1)s}{24\ln 2}\Big)
       + \exp\Big(-\frac{d^k}{24\ln 2 }\Big) \\
&\leq& 6s \exp\Big(-\frac{\a d^k}{512\ln2}\Big)\label{Eq:FinalDeltaBound}.
\end{eqnarray}
Eq.~(\ref{Eq:FinalDeltaBound}) tells us that for any $\ket{\ph} \in S$, in the limit of large $d$, it is overwhelmingly likely that Eq.~(\ref{eq:sufficient}) is satisfied, 
and thus $D^{(X)} \circ E(\ph)$ is close to $\ph$.
However, we would like to bound the probability of error for all states
simultaneously.  To do this, let $ \eta = \frac{\a}{12d^{n-k}}$ and $\cN_S$ be an $\eta$-net for
$S$ with $|\cN_S| \leq (\frac{5}{\eta})^{2s}$.  Then, for $1 \leq k < n$ we find\begin{eqnarray}
	&\,& \Pr \Bigl( \inf_{\ket{\ph} \in S} \min_{i,l} \; 
        \frac{
	\bra{\ph} T_{il} P_l U_i\proj{{\ph}}U_i^\dg P_l T_{il}^\dg \ket{\ph}
	}{
	\<\ph| U_i^\dg P_l U_i |\ph\> 
	}
	 \leq 1-\a \Bigr) 
\nonumber 
\\
\nonumber
	&\leq & \Pr \Bigl(\exists i,l,\ket{\tilde{\ph}} \in \cN_S \mbox{ s.t. }
        {
	\< {\tilde \ph}| T_{il} P_l U_i\proj{{\tilde{\ph}}} U_i^\dg 
	P_l T_{il}^\dg 
		\ket{\tilde{\ph}}
	}
        \leq  (1 -\frac{\alpha}{2})\<\tilde{\ph}| U_i^\dg P_l U_i |\tilde{\ph}\> \Bigr) 
\\ \nonumber
 & & +  \Pr \Bigl(\exists i,l, \ket{\tilde{\ph}} \in \cN_S \mbox{ s.t. } \frac{1}{2 d^{n-k}} > \bra{\tilde{\ph}}U_i^\dg P_l U_i \ket{\tilde{\ph}} \Bigr)
\\
	&\leq& r d^{n-k} \Bigl(\frac{5}{\eta}\Bigr)^{2s} 
	       \Pr \Bigl(\D_{ijl} \geq \frac{\a}{2}\Bigr) \, + r d^{n-k} \Bigl(\frac{5}{\eta}\Bigr)^{2s}\Pr \Bigl ( \frac{1}{2 d^{n-k}} > \bra{\tilde{\ph}}U_i^\dg P_l U_i \ket{\tilde{\ph}}\Bigr). \label{Eq:kLessThann}
\end{eqnarray}
Combining this with Eqs.~(\ref{Eq:denomUB}) and (\ref{Eq:FinalDeltaBound}),
we finally find that 
\begin{eqnarray}
& & \Pr \Bigl( \inf_{\ket{\ph} \in S} \min_{i,l} \; 
        \frac{
	\bra{\ph} T_{il} P_l U_i\proj{{\ph}}U_i^\dg P_l T_{il}^\dg \ket{\ph}
	}{
	\<\ph| U_i^\dg P_l U_i |\ph\> 
	}
	 \leq 1-\a \Bigr)\nonumber\\
&\leq&  6rs d^{n-k} \Bigl(\frac{5}{\eta}\Bigr)^{2s}
        \exp\Big(-\frac{\a d^k}{512\ln2}\Big) 
        +  r d^{n-k} \Bigl(\frac{5}{\eta}\Bigr)^{2s} 
        \exp\Big(-\frac{d^k}{24\ln2}\Big),
\end{eqnarray}
so that if we require
\begin{equation}\label{Eq:BinomBound1}
\binom{n}{k}\Biggl(6rs d^{n-k} \Bigl(\frac{5}{\eta}\Bigr)^{2s}
  \exp\Big(-\frac{\a d^k}{512\ln2}\Big) 
  +  r d^{n-k} \Bigl(\frac{5}{\eta}\Bigr)^{2s} 
  \exp\Big(-\frac{d^k}{24\ln2}\Big)\Biggr) 
\leq \frac{1}{4}
\end{equation}
we guarantee that the probability of any one of the $D^{(X)}$ failing is no
more than $\frac{1}{4}$.  The case of $k=n$ can be analyzed in the 
same way, with the exception that the second term in 
Eq.~(\ref{Eq:kLessThann}) is identically zero, 
resulting in the requirement that
\begin{equation}\label{Eq:BinomBound2}
6rs \Bigl(\frac{5}{\eta}\Bigr)^{2s}\exp\Big(-\frac{\a d^k}{512\ln2}\Big)
\leq \frac{1}{4}.
\end{equation}
Both Eq.~(\ref{Eq:BinomBound1}) and Eq.~(\ref{Eq:BinomBound2}) 
are satisfied if we choose $d$ sufficiently large that
$d^k \geq 48/\d^2$ and, for $k=1$,  $\smfrac{2840(2n+3)}{\d^2} \leq \smfrac{d}{\log d}$,
then choose both $r$ and $s$ according to
\begin{eqnarray}
 \nonumber 
r & = & \left\lceil 
        \frac{32(n+2)^4}{C \e^2} \cdot d^{k-1}\log d \right\rceil \\
s & = & \left\lfloor
        \frac{C\e^2\d^2}{1536(n+2)^4} \cdot \frac{d}{\log d} \right\rfloor.
\label{Eq:RandS2}
\end{eqnarray}
Combining this with our analysis of the probability that a random choice 
of encoding is secure, we also get the second condition
that $d^n > 10(n+2)/\e$. When these requirements are satisfied,  
$E(\cdot)$ and the set of $D^{(X)}(\cdot)$ provide a
$(\d,\e,s,d^n)$ qubit hiding scheme with $(k,n)$ access structure with
probability at least $1/2$.
\begin{comment}
see that if we let
and $d$ sufficiently large that $\frac{d^k}{\log d} \geq \max(144\ln 2(2n+8+\log(\frac{(n+2)^4}{C\e^2})),\frac{36864(2n+3)\ln 2}{\d^2})$, $d \geq \frac{240}{\d^2}$ 
and $d^n >\frac{10(n+2)}{\e} $, we get a secure and accurate $(\d,\e,s,d^n)$
scheme with $(k,n)$ access structure with probability greater than
$\frac{1}{2}$.
\end{comment}

\section{Discussion}

We have shown how to construct multiparty hiding schemes with
threshold access structures for quantum information.  A notable feature 
of these schemes is that in the limit of hiding a large amount
of data, the storage requirement approaches one local physical qubit
per hidden qubit.  That is, the share that each party holds is to 
leading order the same size as the hidden state.  The accuracy and security
parameters incur an overhead that is additive, and therefore
negligible from the point of view of the asymptotic rate.

It seems likely that the threshold schemes presented here can be
concatenated to provide hiding schemes with non-zero asymptotic hiding
rate for arbitrary realizable access structures. (Realizable here
meaning consistent with monotonicity since a superset of an authorized
set must also be an authorized set~\cite{EW02,EW04}.)  
The question of security under concatenation relates to the 
distillability of our encoding states; if a large 
amount of entanglement could be distilled from the encoding states, 
access to some of the encoded 
data could be sacrificed to compromise the security of the rest. 
Luckily, based on the results of~\cite{HLW04}, we suspect that the 
encoding states have at most a small amount of distillable entanglement 
between sub-threshold sets of parties, perhaps even a vanishing amount.
Still, the connection between the theory of multipartite entanglement
and multiparty data hiding is not well understood and deserves further
investigation. At the very least, the states used here have been 
engineered with very extreme properties: small sub-threshold 
distinguishability, high symmetry, high entanglement of formation and
likely low distillable entanglement.

Beyond the question of security under concatenation, it would be
worth knowing whether a composable definition of data hiding could
be formulated and whether the schemes presented here would realize the
definition. In a similar vein, we would like to know the extent
to which the schemes presented here are stable against small amounts
of entanglement shared between the parties; does security fail all
at once or gracefully? Can schemes completely robust against finite
amounts of entanglement be designed? The strongest possible such result 
would be a demonstration that the schemes are secure whenever there is
insufficient entanglement to teleport any local shares.

Moreover, while imperfect security is inevitable
in data hiding (at least in the absence of superselection 
rules~\cite{VC03,KMP03}), 
perfect accuracy is possible, as demonstrated by
\cite{DHT02} and \cite{DLT02}. Is perfect accuracy possible while
simultaneously achieving the rates found in this paper?  Since the
bulk of the technical difficulty in the present paper comes from
proving the existence of sufficiently good, albeit imperfect
decodings, a scheme with perfect decodings could conceivably be
significantly simpler.

Finally, while we have not provided an explicit construction of
the encoding map, it is important to notice that the probability
of $1/2$ in Theorem \ref{thm:main} is arbitrary, and could be
chosen arbitrarily close to one at the expense of increasing
(decreasing) the proportionality constant for $r$ $(s)$ in
Eq.~(\ref{Eq:RandS1}). In this sense, secure and accurate hiding
schemes of the form we present are generic in the limit of large
dimension.  Nevertheless, it would be more satisfying to find explicit,
non-probabilistic choices for the encoding unitaries,
implementable in polynomial time, and still giving secure and
accurate data hiding schemes, as was done for approximate
encryption in \cite{AS04}.

\subsection*{Acknowledgments}
The circuits in Figure \ref{fig:hide} were typeset using the
Q-circuit \LaTeX ~package written by Bryan Eastin and Steven T.
Flammia. The authors would like to thank Carlos Mochon for his optimism
and frequent encouragement during the course of this project.
The authors receive support from the US National Science Foundation
under grant no. EIA-0086038.  PH acknowledges funding from the
Sherman Fairchild Foundation, and DL the Richard
C. Tolman Endowment Fund and the Croucher Foundation.

%%%%%%%%%%%%%%%%%%%%%%%%%%%%%%%%%%%%%%%%%%%%%%%%%%%%%%%%%%%%%%%%%%%%%

\appendix
\section{Success criterion for the PGM with unequal state propabilities}
\label{appendix:PGM}

Using the method of~\cite{HausladenJSWW96}, we will bound 
the probability of error for the PGM when the states
occur with unequal probabilities.  (The symbols used in this
derivation are locally defined.)
Let $\{ |\xi_i\> \}$ be a set of subnormalized states so that $\|
|\xi_i\> \|^2_2$ represents the probability of $|\xi_i\>$, and 
$N = \sum_i  |\xi_i\>\<\xi_i|$. The elements of the PGM are
$M_i = N^{-1/2} \proj{\xi_i} N^{-1/2}$.
We define the matrix
\begin{equation}
	T_{ij} = \< \xi_{i} | \xi_{j} \> \,,
\end{equation} 
which has square root
\begin{equation}
	(\sqrt{T})_{ij} =  \bra{\xi_{i}} N^{-1/2} \ket{\xi_{j}} \,. 
\end{equation}  
In terms of $T$, the probability of the PGM correctly identifying $|\xi_i\>$ 
is 
\begin{equation}\label{Eq:probInTermsofT}
p(i|i) 
= \frac{\bra{\xi_i} M_i \ket{\xi_i}}{\braket{\xi_i}{\xi_i}}
= \frac{\big( \sqrt{T} \big)^2_{ii}} {\<\xi_i | \xi_i \> } \,.
\end{equation}
Applying the inequality $\sqrt{x} \geq \frac{3}{2}x - \frac{1}{2}x^2$
to the matrix $\sqrt{x} =
\frac{\sqrt{T}}{\sqrt{\braket{\xi_{i}}{\xi_{i}}}}$ (which we are
able to do, since the present choices of $\sqrt{x}$ and $x$ are
Hermitian and nonnegative) and noting that the inequality holds for 
the diagonal entries gives
\begin{eqnarray} \label{Eq:squareRootIneq}
	\frac{(\sqrt{T})_{ii}}{\sqrt{\<\xi_i |\xi_i\>}}
 	\; \geq \; \frac{3}{2} \frac{T_{ii}}{\<\xi_i |\xi_i\>} -
	\frac{1}{2} \sum_{j} \frac{|T_{ij}|^2}{|\<\xi_i |\xi_i\> |^2}
	\; = \; 1 - \frac{1}{2} \sum_{j\neq i} 
	\frac{|T_{ij}|^2}{|\<\xi_i |\xi_i\> |^2}
	\; = \; 1 - \frac{1}{2} \sum_{j\neq i} 
	\frac{|\<\xi_i |\xi_{j}\> |^2}{|\<\xi_i |\xi_i\> |^2} \,.
\end{eqnarray}
Combining Eq.~(\ref{Eq:probInTermsofT}) and
Eq.~(\ref{Eq:squareRootIneq}) yields the result
\begin{equation}
p(i|i) \geq 1 - \sum_{j\neq i} 
	\frac{|\<\xi_i |\xi_{j}\> |^2}{|\<\xi_i |\xi_i\> |^2} \,,
\end{equation}
which is equivalent to 
\begin{equation}
p(\neg i | i) \leq \sum_{j\neq i} 
	\frac{|\<\xi_i |\xi_{j}\> |^2}{|\<\xi_i |\xi_i\> |^2} \,.  
\end{equation}

\section{Proof of Lemma \ref{lem:errorControl}} \label{appendix:errorControl}

We would like to show that if $0 < \b,\e \leq 1$, \, $s \leq \b d /128$ \, and $1 \leq k < n$, then
\begin{equation}
\Pr \Biggl(\frac{1}{|\bra{j}U_i^\dg P_l U_i \ket{j}|^2} 
     \sum_{j^\prime \neq j}|\bra{j^\prime}U_{i}^\dg P_l U_i \ket{j}|^2  
     > \beta \Biggr) 
\leq 4s \exp\big(-\frac{\beta d^k}{128\ln2}\big).
\end{equation}
Our argument is somewhat lengthy but completely elementary.
We will make use of the fact that a 
Haar-distributed state in $\CC^{d'}$ can be expressed as
$\frac{|\gamma\>}{\| |\gamma\> \|_2}$ where $|\gamma\>$ is a
$d'$-dimensional vector with all coordinates drawn independently from
$\cN_{\CC}(0,1)$. (Throughout this section, any variable written
as $g_x$, $g_{x}^{y}$ or $g_{xy}^{ab}$, 
for arbitrary values of $x$, $y$, $a$ and $b$, will denote
a random variable drawn from $\cN_{\CC}(0,1)$. They are all chosen
independently.)
To begin, we can express the marginal distributions of the
$U_i\ket{{j^\prime}}$ in terms of complex Gaussians (ignoring
correlations between the vectors $U_i \ket{j_1'}$ and $U_i \ket{j_2'}$):
\begin{equation}\label{Eq:gaussExpans}
U_i\ket{{{j^\prime}}} = 
\frac{1}{\sqrt{\sum_{h=1}^{d^k} \sum_{m=1}^{d^{n-k}}
          |g^{i{j^\prime}}_{hm}|^2 }} 
\sum_{h=1}^{d^k} \sum_{m=1}^{d^{n-k}} 
g^{i{j^\prime}}_{hm} \ket{h}_X \ket{m}_W \,.
\end{equation} 
It will also be useful to note that $\e -\ln(1+\e) \geq \frac{\e}{2}$ 
if $\e \geq 6$, in which case Eq.~(\ref{Eq:epsilonsquared}) can be replaced by
\begin{equation} \label{Eq:epsilon}
\Pr\Big(\frac{1}{N}\sum_{i=1}^N |g_i|^2 \geq (1 + \e)\Big) \leq  \exp\Big(-N \frac{\e}{2 \ln 2}\Big).
\end{equation}
Using Eqs.~(\ref{Eq:epsilonsquared}) and 
(\ref{Eq:epsilon}), the fact that $d^{n-1} \geq 128s$ 
and recalling that $j' = 1,\ldots,s$, we find 
\begin{eqnarray} 
\Pr\Big(\frac{\sum_{i=1}^{{j^\prime}-1}|g_i|^2}{\sum_{i={j^\prime}}^{d^n}|g_i|^2} > \frac{1}{4d}\Big) & \leq & \Pr\Big(\sum_{i=1}^{{j^\prime}-1}|g_i|^2 > \frac{1}{4d}\frac{d^n-{j^\prime}}{2}\Big) + \Pr\Big(\sum_{i={j^\prime}}^{d^n}|g_i|^2 < \frac{d^n-{j^\prime}}{2}\Big)\nonumber\\
& \leq  &\exp\Big(-\frac{d^{n-1}}{32\ln 2}\Big) + \exp\Big(-\frac{d^n}{32 \ln 2}\Big) \leq 2\exp\Big(-\frac{d^{n-1}}{32\ln 2}\Big), \label{Eq:sumOverSumBound}
\end{eqnarray}
which will be useful below.  To complete our task, however, 
we will need to move
beyond the simplest Gaussian approximation to the distribution of the 
$\{U_i\ket{j'}\}$ that takes into account the correlations between
the vectors for different values of $j'$.
%
\begin{comment}
What we need to understand is 
the distribution of
$\{U_i\ket{{j^\prime}}\}$,
recalling that the $\{\ket{{j^\prime}}  \}_{{j^\prime}=1}^{s}$ are mutually orthogonal.  
\end{comment}
%
We can relate $\{U_i\ket{{j^\prime}}\}$ to a collection of independent 
Haar-distributed vectors, $\ket{\psi^{j^\prime}}$, as follows:
\begin{equation} \label{Eq:GramSchmidt}
\ket{\psi^{j^\prime}} = \frac{1}{\sqrt{\sum_{t=1}^{d^n}|g^{j^\prime}_t|^2}}\Biggl(      \sqrt{\sum_{t={j^\prime}}^{d^n}|g^{j^\prime}_t|^2}    U_i\ket{{j^\prime}} + \sum_{t=1}^{{j^\prime}-1}g^{j^\prime}_t U_i\ket{t}\Biggr)
\end{equation}
To see this, notice that $U_i\ket{{j^\prime}}$ is distributed uniformly 
in the orthogonal complement to the span of 
$\{ U_i \ket{t} \}_{t=1}^{{j^\prime}-1}$, 
which means that it can be represented as 
$U_i\ket{{j^\prime}} 
= ({\sum_{t={j^\prime}}^{d^n}|g^{j^\prime}_t|^2})^{-1/2} 
  \sum_{t={j^\prime}}^{d^n} g^{j^\prime}_{t}\ket{b^{j^\prime}_t}$,
where $\{\ket{b^{j^\prime}_t}\}$ forms an orthonormal basis of the 
complement space. Substituting this 
into the expression above shows that $\ket{\psi^{j^\prime}}$ is 
simply a Gaussian state divided by  its norm -- a Haar 
distributed state.  Choosing the $g^{j^\prime}_t$ independently
guarantees that the $\ket{\psi^{j^\prime}}$ will be independent.

Now, without loss of generality, we'll choose $ j = 1$, so that our goal is
to estimate
\begin{equation}
\frac{1}{|\bra{1}U_i^\dg P_l U_i \ket{1}|^2}
\sum_{j^\prime =2}^{s} |\bra{1}U_i^\dg P_l U_i\ket{j^\prime}|^2.
\end{equation}
Inverting Eq.~(\ref{Eq:GramSchmidt}) gives
\begin{equation}
 U_i\ket{{j^\prime}} = \frac{\sqrt{\sum_{t=1}^{d^n}|g^{j^\prime}_t|^2}}{\sqrt{\sum_{t={j^\prime}}^{d^n}|g^{j^\prime}_t|^2}} \ket{\psi^{j^\prime}}  - \frac{1}{\sqrt{\sum_{t={j^\prime}}^{d^n}|g^{j^\prime}_t|^2}}\sum_{t=1}^{{j^\prime}-1}g^{j^\prime}_t U_i\ket{t}
\end{equation}
so that
\begin{equation}
 \bra{1}U_i^\dg P_lU_i\ket{{j^\prime}} = \frac{\sqrt{\sum_{t=1}^{d^n}|g^{j^\prime}_t|^2}}{\sqrt{\sum_{t={j^\prime}}^{d^n}|g^{j^\prime}_t|^2}}\bra{1}U_i^\dg P_l \ket{\psi^{j^\prime}}  - \frac{1}{\sqrt{\sum_{t={j^\prime}}^{d^n}|g^{j^\prime}_t|^2}}\sum_{t=1}^{{j^\prime}-1}g^{j^\prime}_t \bra{1}U_i^\dg P_l U_i\ket{t}.
\end{equation}
Using the inequality $|a + b|^2 \leq 2(|a|^2 + |b|^2)$ then leads to
\begin{eqnarray} \nonumber
 |\bra{1}U_i^\dg P_lU_i\ket{{j^\prime}}|^2 & \leq & 2\Bigl( \frac{{\sum_{t=1}^{d^n}|g^{j^\prime}_t|^2}}{{\sum_{t={j^\prime}}^{d^n}|g^{j^\prime}_t|^2}}|\bra{1}U_i^\dg P_l\ket{\psi^{j^\prime}}|^2 + \frac{1}{{\sum_{t={j^\prime}}^{d^n}|g^{j^\prime}_t|^2}} \Big|\sum_{t=1}^{{j^\prime}-1}g^{j^\prime}_t \bra{1}U_i^\dg P_l U_i\ket{t}\Big|^2  \Bigr)  \\
 & \leq & 2\Bigl[ \Bigl(1 + \frac{{\sum_{t=1}^{{j^\prime}-1}|g^{j^\prime}_t|^2}}{{\sum_{t={j^\prime}}^{d^n}|g^{j^\prime}_t|^2}} \Bigr)|\bra{1}U_i^\dg P_l \ket{\psi^{j^\prime}}|^2 + \frac{\sum_{t=1}^{{j^\prime}-1}|g^{j^\prime}_t|^2}{{\sum_{t={j^\prime}}^{d^n}|g^{j^\prime}_t|^2}}\sum_{t=1}^{{j^\prime}-1}|\bra{1}U_i^\dg P_l U_i\ket{t}|^2  \Bigr].~~\label{Eq:singleIneq}
\end{eqnarray}
Summing over values of ${j^\prime}$ in Eq.~(\ref{Eq:singleIneq}) and using Eq.~(\ref{Eq:sumOverSumBound}) shows that 
\begin{equation}
\Pr\Biggl(\sum_{{j^\prime}=2}^s  |\bra{1}U_i^\dg P_l U_i\ket{{j^\prime}}|^2  >  2\Bigl( \sum_{{j^\prime}=2}^s (1 + \frac{1}{4d})|\bra{1}U_i^\dg P_l \ket{\psi^{j^\prime}}|^2 + \sum_{{j^\prime}=2}^s \frac{1}{4d}\sum_{t=1}^{s}|\bra{1}U_i^\dg P_l U_i\ket{t}|^2  \Bigr)\Biggr) 
%\leq 2s \exp\Bigl(-\frac{d^{n-1}}{128 \ln 2}\Bigr)
\end{equation}
is less than or equal to $2s \exp\Bigl(-\frac{d^{n-1}}{32\ln 2}\Bigr)$,
which implies in turn that
\begin{equation}
\Pr\Biggl(\sum_{{j^\prime}=2}^s  |\bra{1}U_i^\dg P_l U_i\ket{{j^\prime}}|^2  >   4\sum_{{j^\prime}=2}^s |\bra{1}U_i^\dg P_l \ket{\psi^{j^\prime}}|^2 + \frac{s}{2d}\sum_{t=2}^{s}|\bra{1}U_i^\dg P_l U_i\ket{t}|^2   +\frac{s}{2d}|\bra{1}U_i^\dg P_l U_i\ket{1}|^2 \Biggr)
% \leq 2s \exp\Bigl(-\frac{d^{n-1}}{128 \ln 2}\Bigr).
\end{equation}
is bounded above by the same $2s \exp\Bigl(-\frac{d^{n-1}}{32\ln 2}\Bigr)$.
Moving the second sum on the RHS to the LHS and noting $\frac{s}{d} \leq 1$
shows that
\begin{equation}\label{EightIneq}
\Pr \Biggl( \frac{1}{|\bra{1}U_i^\dg P_lU_i\ket{1}|^2}\sum_{{j^\prime}=2}^s  |\bra{1}U_i^\dg P_l U_i\ket{{j^\prime}}|^2  >
   \frac{8}{|\bra{1}U_i^\dg P_lU_i\ket{1}|^2}\sum_{{j^\prime}=2}^s |\bra{1}U_i^\dg P_l \ket{\psi^{j^\prime}}|^2 +\frac{s}{d} \Biggr)
\end{equation}
is again bounded above by $2s \exp \Bigl(-\frac{d^{n-1}}{32\ln 2}\Bigr)$.
Finally, we can upper bound 
\begin{equation}\label{singleSumBeta16}
\Pr \Biggl( \frac{1}{|\bra{1}U_i^\dg P_lU_i\ket{1}|^2}\sum_{{j^\prime}=2}^s |\bra{1}U_i^\dg P_l \ket{\psi^{j^\prime}}|^2 > \frac{\beta}{16} \Biggr)
\end{equation}
by using Fact \ref{fact:2}, Eq.~(\ref{Eq:denomUB}) 
and the estimate leading to Eq.~(\ref{Eq:epsilon}), along with the observation 
that $\ket{\ps^1} = U_i\ket{1}$, 
with the result that the probability in 
Eq.~(\ref{singleSumBeta16}) is less than or equal to
\begin{equation}
\exp\Big(-\frac{\beta d^k}{128\ln2}\Big) + \exp\Big(-\frac{d^k}{24\ln2}\Big).
\end{equation}
Combining this with the bound on Eq.~(\ref{EightIneq}) and noting $\frac{s}{d} \leq \frac{\beta}{2}$ gives the result
\begin{equation*}
\Pr \Biggl( \frac{1}{|\bra{1}U_i^\dg P_lU_i\ket{1}|^2}\sum_{{j^\prime}=2}^s  |\bra{1}U_i^\dg P_l U_i\ket{{j^\prime}}|^2  >  \beta \Biggr) \nonumber\\
\end{equation*}
\begin{equation}
\leq \exp\Big(-\frac{\beta d^k}{128\ln2}\Big) 
     + \exp\Big(-\frac{d^k}{24\ln2}\Big) 
     + 2s \exp \Big(-\frac{d^{n-1}}{32\ln 2}\Big) 
\leq 4s \exp\Big(-\frac{\beta d^k}{128\ln2}\Big).
\end{equation}

%%%%%%%%%%%%%%%%%%%%%%%%%%%%%%%%%%%%%%%%%%%%%%%%%%%%%%%%%%%%%%%%%%%%%

\bibliographystyle{plain}
\bibliography{hide}

\end{document}